\documentstyle[12pt,aaspp4,flushrt]{article}
\doublespace
\begin{document}

\title{A Possible Energy Mechanism for Cosmological Gamma-ray Bursts
}
\author{K.S.Cheng$^{1}$, Y. Lu$^{1,2,3}$}
\affil{ 
$^1$Department of Physics, University of Hong Kong, Pokfulam Road, Hong Kong\\
$^2$Department of physics, Huang Zhong Normal University, 430079, China\\
$^3$ Beijing Astronomical Observatory, Chinese Academy of Sciences, Beijing 100080, China}

\begin{abstract}
We suggest that an extreme Kerr black hole with a mass
 $\sim 10^6M_\odot$, a dimensionless angular momentum $A\sim 1$ and
 a marginal stable orbital radius $r_{ms}\sim 3r_s\sim 10^{12}M_6~cm$
 located in a normal galaxy, may produced a Gamma-ray Burst by capturing and
disrupting a star.  During this period, a transient accretion disk is formed
and a strong transient magnetic field $\sim 2.4\times 10^9M_6^{-1/2}$
Gauss,
 lasting for $r_{ms}/c\sim 30 M_6~s$,
 may be produced in the inner boundary of the accretion disk.
A large amount of rotational energy of the black hole is extracted
and released in the ultra relativistic jet with a bulk Lorentz factor
$\Gamma$ larger
than $10^3$ via Blandford-Znajek process.
 The relativistic jet energy can be converted into $\gamma$-ray radiation
 via internal shock mechanism.
The gamma-ray burst (GRB) duration should be the same as that of the life
 time of the strong transient magnetic field. The maximum number of
 sub-bursts is estimated to be $r_{ms}/h\sim (10 - 10^2)$ because the disk
material is
likely broken into pieces with the size about the thickness of the disk
$h$ at the cusp ($2r_s\le r \le 3r_s$). The shortest rising time of the
burst estimated from this model is
$\sim h/\Gamma c\sim 3\times 10^{-4}\Gamma^{-1}_3(h/r)_{-2}M_6$ s. The model
gamma-ray burst density rate is also estimated.
\end{abstract}


\newpage
\section{Introduction}
Gamma-ray bursts (GRBs) are clearly the "signal" of an extremely energetic
event, which lasts typically a few seconds.
The recent observations of afterglow of GRBs
by BeppoSAX provide strong evidence of the cosmological origin (Metzger et
al. 1997). For example, GRB971214 is the third GRB with a known
optical counterpart. It was detected by the BeppoSAX Gamma-ray Burst
Monitor (Frontera et al. 1997) on Dec 14.97 1997, as a 40s long-structured
GRB and the fluence ($>$ 20KeV) is 10$^{-5}$ erg cm$^{-2}$(Kippen et al.
1997).
GRB971214 was also observed by the All Sky Monitor
on board the X-ray Satellite XTE (Doty 1998). The fluence in the 2-12KeV
band
is estimated to be $(1.8\pm 0.03)\times 10^{-7}\rm erg~cm^{-2}$ (Kulkarni
et al. 1998).
GRB971214 is one of special interesting events
since the burst energy
may exceed $10^{53}$ ergs if its redshift is indeed 3.42 and the
radiation is isotropic (Kulkarni et al 1998).  The observation of
GRB971214
puts serious constrains on the existing theoretical models, especially
on the neutron star merger models (e.g. Narayan et al. 1992)
in which the entire energy
produced during the coalescence is required to release in
gamma-rays in order to explain the observed power of GRB971214.
However the process
of the annihilation of neutrino and anti-neutrino used in this model
may fail to produce the required energy in GRBs (Janka
et al. 1996).  Furthermore this model may also suffer from the
so-called baryon contamination which one has to construct a
special model to avoid (M\'esz\'aros \& Rees 1994).

Before the BeppoSAX no association of GRBs with normal galaxies (Fenimore et
al. 1993)
or Abel cluster (Hurley et al. 1997) has been found.  Recently, the studies
of GRB's afterglows provide the small offsets of GRBs with respect to their
host galaxies and the association of GRBs with dusty regions and
star-formation regions, which favors that the host galaxies of GRB may be
faint galaxies (like a normal galaxy)( Bloom et al. 1999). In fact, at
least, five GRBs have been found to associate with the normal galaxies.
Paczynski
(1998) suggest that this provides strong evidence to support that
 at least some
GRBs are associated with normal galaxies.
 GRBs could be associated with nuclei of normal galaxies harbored
 a massive black hole, whose mass is typically
 $10^5M_\odot\le M\le 10^6M_\odot$(Roland et al. 1994). It has also been
argued that GRBs might be associated with some types of quasars, such as
radio quiet quasars (Scharted et al 1997) and metal-rich quasars
(Cheng et al. 1997; Cheng \& Wang 1999). There are two classes of quasars,
i.e. inactive
quasars and active quasars, in nature (Rees 1990). The host galaxies of
inactive (quiescent) quasars may appear as normal galaxies (Burderi , King
\& Szuszkiewicz 1998). With the quasars
luminosity function one can obtain the mean density of the observed
active quasar
in the Universe about $n_{QSO}=10^2$ Gpc$^{-3}$ (Woltjer 1990; Osterbrock
1991).  The time of the active phase of a quasar is
$t_{QSO}\sim 10^8$ yr.  A similar result can also be
obtained independently from the quasar statistics for a wide range of
redshifts (Phinney 1992). Some observations seem to support this
simple estimation (Artymowicz et al. 1993).  This typical life time
is much shorter than Hubble time $10^{10}$ yr.  It implies that
the inactive quasar, where the central engine was
extinct due to starving of fuel (Rees 1990), should have a mean number
density
$n_{IQSO}=10^4$ Gpc$^{-3}$. Rees (1984)
pointed out that the massive black hole is one of the favored sources
for powering the quasar if an inactive quasar has a central massive
black hole. Lynden-Bell (1969) first suggested that inactive quasar are in
the form of massive underfed black holes, which may be present in
the dense nuclei of normal galaxies, and
Cannizzo (1990) has generalized that most "normal"
galaxies may also harbor black holes in their centers. In fact, the recent
detection of a $\gamma$-ray flux along the direction of the Galactic center
by GRGET on the broad Compton GRO suggested that these $\gamma$-rays may be
originated close to the massive black hole (Mastichiads \& Ozernoy 1994). By
a more careful treatment of the physics of p-p scattering, Markoff, Melia \&
Sarcevic (1997) suggest that the black hole of mass $10^6M_\odot$ may exist
in the Galactic center and contributing to these high-energy emission.

In this paper, we investigate the possibility of
inactive quasars, which have harbored rapidly spinning black
holes embedded in a dense star cluster
as the hosts of $\gamma$-ray bursts. When a star is tidally captured and
disrupted
by the black hole, part of the stellar matter will be swallowed
by the black hole.  A transient accretion disk/torus
then can be formed and the inactive quasar will be activated
(Rees 1988; Cannizzo 1990; Sanders 1984).  During the active phase of
the quasar, the rotational energy of the massive black hole can be extracted
and quickly converted into the bulk kinetic energy of the magnetically
driven
outflow by Blandford-Znajek process (Blandford \& Znajek 1977).
A relativistic fireball
wind is formed and results in a GRB (e.g.Rees \& M\'esz\'aros 1994). This
paper is organized as following. In section 2, we describe how the black
hole
captures a main sequence star and forms a transient accretion
disk. The capture rate is also estimated. In section 3, we discuss under
what conditions a strong energy outburst via the Blanford-Znajek mechanism
can occur and the various time scales associated with GRBs in this model are
estimated. The GRB burst rate of this mechanism and a brief discussion are
presented in section 4.


\section{Tidal disruption and formation of accretion disk}

The rate at which a massive black hole in a dense star cluster tidally
disrupts and swallows stars has been studied by many authors (e.g.Hills
1975;
Bahcall et al. 1976; Lightman et al. 1977). An important step was the
realization by Frank and Rees (1976) and Lightman et al.(1977) that the
problem was essentially one of loss-cone diffusion-diffusion in angular
momentum rather than energy. The maximum swallowing rate may occur for
stars not near the black hole but at some relatively large distance from
the black hole, at the so-called critical radius where the root-mean-square
angular diffusion of stellar velocity vectors (due to two-body encounters)
over one orbital period is comparable to the angular cross section of the
black hole (the loss cone angle).
The ideas presented in this section closely parallel to those discussed by
Rees (1988).
The only difference is that Rees (1988) has argued that the disrupted
star forms a thick hot ring at the tidal radius with
quite short accreting time scale,
but we shall discuss the case of thin accretion disk formed by the disrupted
star
as suggested by Sanders (1984), which is also possible.

When a star whose trajectory happens to be sufficiently
close to the massive black hole, the star would be captured and eventually
tidally disrupted . After a dynamics time scale ( orbital time scale),
the debris of a tidally disrupted star will form a transient accretion disk
around the massive
black hole with a radius typically comparable to the tidal radius (Rees
1988).
On a time scale  determined by the rate
at which angular momentum can be transferred by a presumed turbulent
viscosity, the torus of debris will be swallowed by the massive black hole.
The inactive quasar with the massive black hole is activated only over this
swallowing time scale,which is $\sim$ 1 yr for a thick hot ring (Rees 1988)
and is $\sim 10^2$ yrs for a thin cool disk.
The following physical quantities are relevant to this problem:

(a) The maximum tidal disruption distance for a captured star

If a Newtonian approximation for the gravitational field is used,
the tidal disruption radius is (Sanders 1989)
\begin{equation}
R_T= (\frac{6M_{BH}}{\pi\bar{\rho_*}})^{1/3}
=1.4\times 10^{13}(\frac{M_*}{M_\odot})^{-1/3}
(\frac{M_{BH}}{10^6M_\odot})^{1/3}(\frac{R_*}{R_\odot})~~{\mbox{cm}}\;\;,
\end{equation}
where $\bar{\rho_*}$ is the mean mass density
of matter in the star, $R_*$ and $M_*$ are the
radius and mass of the incoming star respectively.
Notice that the ratio of $R_T$ to
the Schwarzschild radius ( $r_s=\frac{2GM_{BH}}{c^2}$) is
about $\sim 50(M_6)^{-2/3}$
($M_6=M_{BH}/10^6M_\odot$). So the Newtonian approximation is indeed a
reasonable one if $M_6 \sim 1$. We choose the typical mass $\sim
10^6M_\odot$ of a black hole because the afterglow of GRB observations
suggest that the host galaxies are likely faint normal galaxies(Bloom et
al.1999, Menten et al. 1997; Marfoff Melia \& Sarcevic 1997; Astichiadis \&
Ozernoy 1994; Roland et al. 1994).

(b) The tidal capture and disruption rate

We note that the schwarzschild radius increases more rapidly with
black hole  mass than that of the tidal radius. Therefore, a star captured
by
a black hole with mass $M_{BH}\sim 3\times 10^8 M_\odot$ will be swallowed
before being tidally disrupted (Hills 1975).
We can reasonably estimate the capture and tidally disrupted rate
$\dot{N}_c$ in terms of the density and velocities of the surrounding stars.
For a star cluster scenario of our model we assume that the constituent
stars are main sequence stars of identical proper masses.
Lacy \& Townes (1980) discussed that the internal dispersion velocity $v_s$
of these stars is $\sim 100 km s^{-1}$ and their density $n \sim 10^2$ in
the inner one parsec core of the galactic nucleus. Based on Cohn et al. 1978
about the captured rate, we obtain:
\begin{equation}
\dot{N}_c \sim 10^{-7}\left(\frac{M_{BH}}{10^6}\right)^{2.33}
\left(\frac{n}{1\times 10^2\rm{pc}^{-3}}\right )^{1.60}
\left (\frac{v_s}{100~\rm
km~s^{-1}}\right )^{-5.76}(\frac{M_*}{M_\odot})^{1.06}
(\frac{R_*}{R_\odot})^{0.4} ~~\rm{yr^{-1}}.
\end{equation}
This could be an underestimate, since the observed stellar density rise
more rapidly than that used in Cohn's fully relaxed
stellar cusps (Cannizzo 1990). However, no serious modification
to above simple estimate $\dot{N}_c$ shall appear (Rees 1988).

(c)The evolution of the transient accretion disk

Cannizzo (1990) pointed out that two processes will supply mass to the
central
black hole after the tidal disruption of a star occurs: (1) the stream of
stellar matter
strung out in far-ranging orbits, and (2) mass loss from the inner edge of
the
accretion disk. Rees (1988, 1990) discussed the former process extensively
and shows that, the infall rate declines as $t^{-5/3}$, and the evolution
of the swallowing rate (accretion rate) is given by
\begin{equation}
\dot{M}=25M_6^{-1/2}(t/t_D)^{-5/3}M_{\odot}~~\rm yr^{-1}~~,
\end{equation}
The Eddington accretion rate of a massive black hole  $10^6M_\odot$ allowed
is
\begin{eqnarray}
\dot{M}_{Edd}=10^{-2}M_6 ~~M_\odot yr^{-1};
\end{eqnarray}
where $t_D$ is the dynamics or orbit time scale. For a central black hole of
$10^6M_\odot$, the orbital period is
given by Sanders (1984)
\begin{equation}
t_D=\Omega^{-1}=(\frac{GM_{BH}}{R_T^3})^{-1/2}
=4.54 \times 10^2 M_6^{1/2}~~\rm s~~.
\end{equation}
from the Eq(3) and Eq(4), one can get that the accretion rate of the disk
will equal $\dot{M}_{Edd}$ at $\sim 0.5$ days and so most effective
radiation would be concentrated on the dynamics time.
Owing to the continued mass supply at late time, the effective falloff of
$\dot{M}$ with t may be less steeper than $t^{-5/3}$. Cannizzo (1990) has
discussed the case of the late time evolution of $\dot{M}$ in which the
accretion disk supply rate varies as $t^{-1.2}$.


\section{Energy mechanism of GRB}

Once a transient accretion disk surrounding the rapidly spinning black hole
is
formed as a result of the processes described in previous section,
there could be an ordered poloidal field threading the black
hole,  associated with a current ring in the disk.
This ordered field can extract energy via the Blandford-Znajek
process, creating magnetically driven outflows (jets) along
the rotation axis (Blandford \& Znajek 1977).
Irrespective of the detailed field structure, any magnetically driven
outflows along the rotation axis are less loaded with baryons than in
other directions (M\'esz\'aros 1997).
The role of the disk is mainly to anchor the magnetic field:
the power comes from the rapidly spinning black hole itself, whose
rotational energy can be $\sim 10^{60}M_6$ ergs.
However, while an adequate energy source for GRB is available,
like those discussed in previous section, two major problems arise.
Namely, how the rotation energy of the black hole
can be converted into an adequate photon flux in the
right energy range, and what causes these photons to appear in bursts of
$\sim 1-100~s$ duration. Two classes of fireball models
,which provide different explanations for the duration and variability of
GRB, are generally used.
The first class is called external shock models (e.g.
M\'esz\'aros \& Rees 1993) which
are caused by the interaction (collision) between the fireball eject and
the surrounding medium. The typical duration of a GRB is then given by the
Doppler delayed arrival times of the emission from the two boundaries of the
eject shell, or from the delay between different surface elements within
the light cone. The second class is called internal shock models (e.g.Rees
\&
M\'esz\'aros 1994; Paczynski \& Xu 1994) which
relates the shocks to inhomogeneities
within the relativistic outflow, e.g. catching up of faster portions with
slower portions of the flow. The duration of these shocks is likely to be
given by the intrinsic duration of the energy release.
According to the internal shocks model of Rees \& M\'esz\'aros (1994),
relative motions within the outflow material will be relativistic and lead
to internal shocks resulting
in relativistic fireball winds as well as triggering a GRB on the time of
a few seconds if the mean Lorentz factor $\Gamma$ fluctuates by a factor of
$\sim 2$ around its mean value.
\subsection{The energy of GRBs}
The tidally disrupted star by a massive black hole is assumed
to form a thin disk. According to the standard accretion
disk theory (Shakura $\&$ Sunyaev 1973; Novikov \& Throne 1973), the maximum
pressure in the disk can be approximated by
\begin{eqnarray}
p_{d.max}=
\left\{
\begin{array}{ll}
2\times 10^{10}(\alpha M_6)^{-1}\Re_1  ~~~
{\mbox{dyne~cm$^{-2}$}}~~~~
\dot{m}>\dot{m}_c \\
8.013\times 10^{12}(\alpha M_6)^{-9/10}\dot{m}^{4/5}\Re_2
~~~{\mbox{dyne~cm$^{-2}$}} ~~~~ \dot{m}<\dot{m}_c \;\;,
\label{P_dmax}
\end{array}
\right.
\end{eqnarray}
and
\begin{equation}
\dot{m}_c=5.98\times 10^{-3}(\alpha
M_6)^{-1/8}(\frac{\tilde{R}_1}{\tilde{R}_2})^{5/4}\;\;.
\label{dotm_c}
\end{equation}
The density $n$ of the disk at $\dot{m}>\dot{m}_c$ is
\begin{eqnarray}
n &=& 2.79\times 10^{20}{\alpha M_6}^{-1}~~{\mbox{cm$^{-3}$}}~~;
\end{eqnarray}
where $\alpha$ is a viscous parameter($0<\alpha<1$),
$\dot{m}=\dot{M}/\dot{M}_{Edd}$,
$\dot{M}_{Edd}=L_{Edd}/c^2=1.3\times 10^{44} M_6/c^2 ~erg s^{-1}$.
$\Re_1\sim \tilde{r}_{ms}^{-3/2}, \Re_2\sim\tilde{r}_{ms}^{-51/20}$,
$\tilde{r}_{ms}=\frac{2r_{ms}}{r_s}$ and $r_{ms}$ is the inner most stable
orbit.
The upper formula in equation (6) corresponds to the case of the maximum
pressure dominated by radiation, while the lower formula in equation (6)
corresponds to the case of the maximum pressure dominated by gas. So,
for high accretion rates, the maximum pressure does not depend on the
accretion rate. It can be checked by noting that disk pressure is
$\propto \dot{m}/\alpha M(h/r)$ and
that $h/r\propto\dot{m}$ for radiation pressure dominated disk, where $h$ is
the height of an accretion disk.
For $\dot{m} > \dot{m_c}$, the maximum pressure in innermost parts of
the geometrically thin accretion disk is dominated by the radiation
pressure.
If magnetic viscosity is dominant, then
\begin{equation}
p_{d.max}=B_{eq}^2/8\pi\approx p_r >> p_g~~,
\end{equation}
where $p_r$ and $p_g$ are radiation pressure and gas pressure
respectively. From Eq.(6)and Eq.(9), we obtain that $B_{eq}\sim
10^5M_6^{-1/2}$ Gauss for the case of high accretion rate.

10
With the evolution of the disk,
when the accreted matter reaches the inner region of the disk (inside
$3r_s$),
a bursting growth of magnetic field occurs. It
 is possible that the flux-freezing in the differential rotating
disk can cause the seed and/or generated magnetic field to wrap up tightly,
becoming highly sheared and predominantly azimuthal in orientation: a
manifestation of the dynamo effect elucidated by Parker (1970). Such
poloidal
fields can extract angular momentum from the disk, enabling efficient
accretion of disk plasmas onto black holes. Haswell et al (1992) estimate
that the growth time scale of the magnetic field is
$\Delta t_g\sim \left(\frac{2}{9}\right)^{1/2}
\left(\frac{L_B}{V_A\sqrt{R_m}}\right)^{1/4}
\left(\frac{r_s}{c}\right)^{3/4}$.
The amplified magnetic field is limited by the gravitational energy
\begin{equation}
B_{max}\sim \frac{c}{V_A}\left(\frac{L_B}{r_s}\right)^{1/2}B_{eq}\sim
2.4\times 10^9M_6^{-1/2} ~~\rm{Gauss};
\end{equation}
where $V_A=B_{eq}/\sqrt{4\pi\rho}, \rho=nm_p$ is the Alfven velocity
,$m_p=1.67\times 10^{-24}g$, $R_m=10^{10}$ is the
magnetic Reynolds number (Haswell 1992) and $L_B$ ($\sim r_{ms}$) is the
radial size of the magnetic
structure. For simplicity, we have chosen $\alpha=1$. Assuming $M_{BH}\sim
10^6 M_\odot$ and the black hole
accreting at the Eddington
rate, $\Delta t_g\sim 31~s$ and
$B_{max} \sim 2.4\times 10^9$ Gauss.
According to Lynden-Bell \& Pringle (1974),
when the magnetic field is so strong that it dominates over the material
pressure, all
the material within the inner boundary region will fall onto the central
black
hole rapidly. It is noted that no viscous energy losses could take place at
$r<r_{ms}=3r_s$ (Shakura \& Sunyaev 1973). Therefore, the debris with mass
$\Delta M$ in the disk
\begin{eqnarray}
\Delta M \sim 2\pi r_{ms}h^2m_pn\sim 2\times
10^{29}M_6^2\left(\frac{h}{r_{ms}}\right)^2_{-2}~~\rm{gm};
\end{eqnarray}
where
$\left(\frac{h}{r_{ms}}\right)_{-2}=\left(\frac{h}{r_{ms}}\right)/10^{-2}$
will fall onto the black hole on a free fall time scale $t_{ff}$ given by
\begin{equation}
t_{ff} \sim r_{ms}/c \sim 28M_6  ~~\rm{s}
\end{equation}
This time scale is very close to the growth time scale $\Delta t_g$ of
$B_{max}$.  We want to point out that $B_{max}\sim 2.4\times 10^9$ G can
only occur no more than once during the entire accretion process.
It is because it takes a viscous time scale
$\sim (r/h)^2t_D\sim 4.5\times 10^6(r/h)^2_2M_6^{1/2}$ s
 to replenish the mass loss, however, the accretion rate ({\it cf.}
Eq.(3)) becomes several orders of magnitude lower than the Eddington rate.
Since in such disks no viscous energy losses are taking
place at $r < r_{ms}$, the process of the massive black hole
swallowing the debris must be completed by dissipating its own rotation
energy ($Pdt$) and angular momentum ($Pdt/\Omega_F$) via the
Blandford-Znajek
process. The total Blandford-Znajek power is given by ({\it cf}. Lee, Wijers
\& Brown 1999)
\begin{equation}
P=1.7\times 10^{50}A^2f(A)M_6^2(\frac{B_n}{10^9 gauss})^2~~ergs^{-1};
\end{equation}
where $B_n$ is the intensity of a magnetic field component normal to the
black
hole horizon, $f(A)=0 (A=0) \rightarrow 1.14(A=1)$, which is a constant for
a given $A$.
$A$ is the dimensionless angular momentum
($A=\frac{2GM_{BH}\Omega_h}{c^3}$), and $\Omega_h$ is the angular velocity
of the black hole.  So, the maximum power extracted by the Blandford-Znajek
process is
\begin{eqnarray}
 P_{max}&\approx& 10^{51}f(A)A^2M_6~~ erg s^{-1};
\end{eqnarray}
where $B_n=B_{max}$ is used and  the maximum GRB energy is given by
\begin{equation}
(E_b)_{max} \approx 3\times 10^{52}A^2f(A)M_6^2~~ \rm{ergs}.
\end{equation}
We should note that the magnetic field generated during this process will
accelerate plasma toward both polar directions by the $J\times B$ force,
and the accelerated plasma form bipolar relativistic jets (magnetically
driven outflow) collimated by the magnetic force. This magnetic
mechanism is the same as that proposed for AGN jets (Lovelace 1976;
Blandford \& Payne 1982; Pelletier
et al. 1996; Meier et al. 1997), and similar to that suggested by Dai \& Lu
(1998a,b; 1999) who considered a postburst shock renewed by the
electromagnetic waves radiated from a puslar. According to the simulations
of relativistic
jets driven by non-steady accretion of magnetized disk of Koide et al.
(1998),
there exist two-layered shell structure jets consisting of a fast jet in the
inner part and a slow jet in the outer part, both of which are collimated by
the global poloidal magnetic field penetrating the disk. When the faster
portions catch up with the slower portions of the outflow, the internal
shock is produced which eventually triggers GRB (Rees \& M\'erz\'aros 1994).
The typical Lorentz factor of this relativisitic jet can be estimated as
$\Gamma_{max}\sim (E_b)_{max}/c^2f\Delta M$, where $\Delta M$ is the mass of
the debris falling onto the black hole and $f$ is the fraction of the debris
material rejected by the relativistic jet. Since the accretion flow near the
black hole should  be isotropic, then $f\sim \Delta \Omega/4\pi$, where
$\Delta \Omega$ is the solid angle of the jet. According to the observations
of Active Galactic Nuclei (AGN), which should also contain a massive black
hole, the opening angle of the jet $\sim 15^o - 30^o$. So $f$ should be
approximately $\sim 0.1-0.01$, the typical Lorentz factor of the jet
material should be
\begin{eqnarray}
\Gamma_{max}\sim 10^3f_{0.1}^{-1}\left(\frac{h}{r_{ms}}\right)^2_{-2};
\end{eqnarray}
where $(h/r_{ms})_{-2}=(h/r_{ms})/10^{-2}$.

\subsection{The Variabilities of GRBs}

 There is unlikely any relativistic motion between the host galaxy and the
Earth, the duration of the burst must be
\begin{eqnarray}
\Delta t_{busrt}\sim M_{max}(\Delta t_g, t_{ff})\sim 30M_6~s;
\end{eqnarray}
In addition, there are at least two time variabilities associated with this
type of GRB model. First, since the debris material are likely broken into
clumps/blobs which should have a characteristic dimension $h$, by the strong
magnetic field $B_{max}$ before they are dragged onto the black hole. These
blobs are expected to be wrapped up by magnetic field lines, so they can
maintain their size for a diffusion time scale $t_{diff}\sim h^2/r_lc$,
where $r_l$ is the Larmor radius. Since protons inside the blobs shall not
have energy larger than $\Gamma m_pc^2$ so $
t_{diff}>10^{10}(h/r_{ms})^2_{-2}\Gamma_3^{-1} B_9 s$ . Therefore the burst
should consist of some sub-bursts and the maximum number of sub-bursts is
estimated as $r_{ms}/h$. It has been shown (Abramowicz \& Zurek 1981) that
accretion flow close to $r=r_{ms}=3r_s$ is transonic. Due to general
relativistic effects, close to black hole, the relation of it's thickness
$h$ with radii $r$ in the range of $r_{ms}\ge r\ge 2r_s$ is (Abramowicz
1985)
\begin{eqnarray}
\frac{h}{r} &\sim& 10^{-2}\sqrt{\beta}/x ;
\end{eqnarray}
where $x$ is a parameter depending on the mass of the black and the
accretion rate, which are $x\sim 1$ for stellar black holes and $x\sim 0.1$
for massive black holes. $\beta$ is the ratio of the gas pressure to the
total pressure of the disk and with $1\ge \beta\ge 10^{-4}$. Then, the
number of sub-bursts is about
\begin{eqnarray}
N_{max}=r_{ms}/h= 10\beta^{1/2}_{-4}x_{0.1}^{-1}   ~~.
\end{eqnarray}
where $\beta_{-4}=\beta/10^{-4}, x_{0.1}=x/0.1$. The typical number of
sub-bursts of a GRB is about $10 - 10^2$ (Sari \& Piran 1997).
Another fine time scale in this model is the shortest rising time
 of the burst. Since each of these blobs is moving with
a Lorentz factor $\Gamma_{max}$ and a dimension $h$ in the comoving frame.
 When these blobs convert their kinetic energy into radiation,
the rising time scale of the the radiation should be
\begin{eqnarray}
\Delta t_{rise}\sim \frac{h}{\Gamma c}\sim 3\times
10^{-4}\Gamma_3^{-1}(h/r_{ms})_{-2}M_6 ~~s.
\end{eqnarray}
where $\Gamma_3=\Gamma/10^3$. Here, we have used the fact that the size of
blobs does not evolve in a
 time scale of $10^{10}s$ because they are wrapped up by the magnetic
 field lines ({\it cf.} previous section).


\subsection{GRBs with possible host galaxies}
 Severals strong GRBs detected by BATSE and BeppoSAX are suggested to
associate with distance normal galaxies with redshift $z>1$ (except
GRB980425, and GRB970228), whose observed features are listed in Table 1.
Five of these GRBs show complex temporal structure, in particular their
variability time scale is significantly shorter than the duration. In fact,
most GRBs show sub-burst structure and the number of sub-burst is $\sim
10 -10^2$ and the rising time of the burst/sub-burst could be shoter than
millisecond (Greiner 1998). We use the properties of GRB971214 to illustrate
our model.
The fluence of the recently observed GRB971214 (Kulkarni et al. 1998)
corresponds to
\begin{eqnarray}
E_\gamma=(\frac{\delta \Omega_\gamma}{4\pi})10^{53.5} ~~\rm{ergs};
\end{eqnarray}
 which is consistent with $(E_b)_{max}$ if a beaming
factor $\frac{\delta\Omega_\gamma}{4\pi}$ is $\sim 0.1$.
Taking $M=1.3\times 10^6 M_\odot$,
 then we can fit the duration of GRB971214 $\sim 40 ~s$ ({\it cf.}
Eq.(17)). The number of sub-burst of GRB971214 is $\sim r_{ms}/h \sim 10$ if
$\beta\sim 10^{-4}$ , which is consistent with the fact that the radiation
pressure dominates the gas pressure at high accretion rate. The shortest
rising timescale
$t_{rise}\sim  3\times 10^{-3}$ s.


\section{Discussion}

We have proposed a new GRB model which is associated
with an inactive quasar/normal galaxies harboring a spinning massive black
hole with mass $10^6M_\odot$ embedded a star cluster, a speculative case is
GRB971214. We have discussed the rising time, sub-burst time, duration and
energy mechanism of GRBs.
According to the capture rate by the tidal force of the black hole
(see Eq.(2)) and the number density of inactive quasar with spinning
black hole (see $n_{IQSO}$), we  also can estimate the rate of an inactive
quasar which can be converted into
an active quasar by tidal disrupted star is
$\rho=\dot{N}_c n_{IQSO}=1.55\times 10^{-6}~~~\rm Gpc^{-3}yr^{-1}$.
If GRBs are only associated with inactive quasar harboring rapidly massive
black
hole, then we can obtain the density rate of burst events,
$\rho_{GRB}=\rho=1.55\times 10^{-6}~~~\rm Gpc^{-3}yr^{-1}$.
This estimation is far lower than the observation results
($22.3\rm Gpc^{-3}yr^{-1}$, Fenimore et al. 1993).

According to recent observations, at least five GRBs
are associated with normal galaxies, which may contain
a black hole with mass $ \sim 10^6 M_{\odot}$ like that in our
Galaxy. Assuming other normal galaxies having similar
properties of our Galaxy, Eq(2) gives the capture rate
$\dot{N}_c \sim 10^{-7} yr^{-1}$. The density of normal galaxies is
$n_{galaxy}
\sim 10^9 Gpc^{-3}$. Combining this result with the capture
rate in normal galaxies, the GRB density rate is
\begin{eqnarray}
\rho_{GRB}&=&n_{galaxy} \dot{N}_c\nonumber\\
&=&10^2 Gpc^{-3} yr^{-1}.
\end{eqnarray}
 If the beaming factor $f$ is included, the GRB rate will increase by a
factor $f^{-1/2}$ because we are able to observe farther GRBs, assuming the
minimum detectable energy flux and the energy release of GRBs are constants.

Although our model resembles to other GRB models
(e.g. microquasar model by Paczynski 1998;
failed supernova model by MacFadyen \& Woosley 1998; Popham,
Woosley \& Fryer 1998) in the way
that both models suggest that the burst energy come from the Kerr black hole
via Blandford-Znajek mechanism. However, our model differs
from their models in the following aspects.
(1)Our accretion disk is formed transiently after capturing a
main sequence star. On the other hand, the disk in other models
is formed by the mass coming from either the progenitor
or ejected during the collapse process of merger.
(2)Since the origin of
the burst is so different, the estimation of burst rate is completely
different. Other models, the burst rate relates either to the merger
rate or the rate of failed
supernova, ours relates to the capture rate.
(3)The accretion rate required by our model is the Eddington rate but in
other models are always
super-Eddington rate( usually 5 to 7 orders of Eddington rate. It still
requires more detailed study.)
(4)The main energy release in our model is the rotation energy of the
black hole NOT the accretion energy( this is exactly why we need to
use the Blandford-Znajek Mechanism which can extract the rotation energy
of the black hole in an extremely efficient way).
(5)Since the GRB source is a normal galaxy (inactive quasar) which
is activated by capturing a main sequence star.
The accretion disk in our model therefore is a transient accretion disk
and the gamma-ray burst is triggered by the disk instability.
The identification of galaxy to GRBs does not conflict with our
model.

Although our mechanism is also similar to that of quasar/ AGN
model (e.g. Blandford-Znajek 1977), there are
two major differences. First, the maximum outburst
power ({\it cf.} Eq.(14)) is much larger than that of the the flare
state of quasar because
 a transient magnetic field $B_{max}$ is expected to occur in GRB
 models. This exponential increase of the magnetic field
from $B_{eq}$ to $B_{max}$ is due to an instability in the
 inner region of the disk, e.g. due to mass accumulation
resulting in the increase of the viscosity and local accretion rate and
 developing an instability eventually. Second, the flare state of
quasar/AGNs
could be repeated in time
scales from days to years.
  Our model does not predict repeated GRBs because:
 After the GRB, the disk losts a mass ring ({\it cf.} Eq(11)). It takes a
viscous time scale $\sim 4\times 10^6 s$ to replenish the mass. The
accretion rate of the disk will decrease to $\sim 10^{-3}$ of the Eddington
rate (${\it cf.}$ Eq.(3)),
 then the radiation pressure $p_{max}$ (${\it cf}.$ the lower part of
Eq.(6)) is too low to produce enough magnetic field to trigger GRB again.

Acknowledgments:\\
We thank an anonymous referee for important suggestions and Z.G.Dai,
J.H.Fan, T.Lu, J.M.Wang, D.M.Wei and L.Zhang for very useful discussions and
comments. This work is partially supported by a RGC grant of Hong
Kong Government and the National Natural Science Foundation of China.

\clearpage
\begin{table}
\caption{Observed features of five GRBs with possible host galaxies.}
\begin{tabular}{lllll}\\\hline
Burst Name & $\Delta t$ (sec) & z & $E_\gamma$ ~(erg) & $N_{sub}$ \\\hline
GBR970228 & 80  & $\sim 0.695$ & $\sim 5.2\times 10^{51}$ & $\sim 4$\\
GRB971214 & 40  & $\sim 3.418$ & $\sim 3\times 10^{53}$ & $\sim 6$ \\
GRB980425 & 20  & $\sim 0.0085$ & $\sim 5\times 10^{47}$ &$\sim7$\\\
GRB980613 & 20  & $\sim 1.0964$ & $\sim 5.2\times 10^{51}$ &$\sim 1$\\
GRB990123 & 100 & $\sim 1.6$  & $\sim 3.4\times 10^{54}$ & $\sim 8$
\\\hline
\end{tabular}
\\\

$\Delta t$ is duration, $z$ is redshift, $E_{\gamma}$ is
the energy of GRB assuming isotropic emission \\
and $N_{sub}$ is the sub-burst number of GRB, respectively.\\
References:\\
(1) GRB970228 (Costa \& Feroci et al. 1997, IAU Cir. No.
6572);\\
(2) GRB971214 (Kippen et al. 1998, IAU Cir. No. 6789);\\
(3) GRB980425 (Kippen et al., 1998,GCN, No., 110); \\
(4) GRB980613 (Wood \& Kippen et al. 1998,GCN, No. 112);\\
(5) GRB990123 ( Kippen et al. 1999, GCN, No. 224)
\end{table}

\end{document}